# High-precision realization of robust quantum anomalous Hall state in a hard ferromagnetic topological insulator

Cui-Zu Chang,[1] Weiwei Zhao,[2] Duk Y. Kim,[2] Haijun Zhang,[3] Badih A. Assaf,[4] Don Heiman,[4] Shou-Cheng Zhang,[3] Chaoxing Liu,[2] Moses H. W. Chan,[2] and Jagadeesh S. Moodera[1,5]

[1]*Francis Bitter Magnet Lab, Massachusetts Institute of Technology, Cambridge, MA 02139, USA*

[2]*The Center for Nanoscale Science and Department of Physics, The Pennsylvania State University, University Park, PA 16802-6300, USA*

[3]*Department of Physics, Stanford University, Stanford, CA 94305-4045, USA*

[4]*Department of Physics, Northeastern University, Boston, MA 02115, USA*

[5]*Department of Physics, Massachusetts Institute of Technology, Cambridge, MA 02139, USA*

Corresponding authors: czchang@mit.edu(C. Z. C.); wzhao@phys.psu.edu (W. Z.) and moodera@mit.edu  (J. S. M.)

## Abstract

The discovery of the quantum Hall (QH) effect led to the realization of a topological electronic state with dissipationless currents circulating in one direction along the edge of a two dimensional electron layer under a strong magnetic field.[1,2] The quantum anomalous Hall (QAH) effect shares a similar physical phenomenon as the QH effect, whereas its physical origin relies on the intrinsic spin-orbit coupling and ferromagnetism.[3-16] Here we report the experimental observation of the QAH state in V-doped $(Bi,Sb)_2Te_3$ films with the zero-field longitudinal resistance down to $0.00013\pm0.00007 h/e^2$ (~$3.35\pm1.76\Omega$), Hall conductance reaching $0.9998\pm0.0006 e^2/h$ and the Hall angle becoming as high as $89.993\pm0.004$ °at $T$=25mK. Further advantage of this system comes from the fact that it is a hard ferromagnet with a large coercive field ($H_c$>1.0T) and a relative high Curie **temperature. This realization of robust QAH state in hard FMTIs is a major step towards dissipationless electronic applications without external fields.**

The quantum anomalous Hall (QAH) effect describes the dissipationless quantized Hall transport in ferromagnetic materials in the absence of external magnetic fields.[3-13] The realization of the QAH effect in realistic materials requires two conditions: ferromagnetic insulating materials and topologically non-trivial electronic band structures.[5,12,13] It has been proposed that these two conditions can be satisfied by introducing ferromagnetism into topological insulators (TIs).[12,13] Ferromagnetism can be achieved by doping TIs with transition metal atoms, such as Cr, V and Mn.[17,18,19] Indeed, the QAH effect has been reported in Cr-doped $(Bi,Sb)_2Te_3$ system.[14,15,16] Among the various transition metal atoms, V-doped $Sb_2Te_3$ exhibits the most stable ferromagnetism with a high Curie temperature ($T_C$).[17,18] This fact motivated us to explore the QAH effect in V-doped TI thin films. We succeeded in realizing in this system a robust QAH state with high-precision, making such devices more amenable for metrology and spintronics applications.

Epitaxial thin films with various concentrations of V-doped $Sb_2Te_3$ were prepared by coevaporation in a molecular beam epitaxy (MBE) system. Excellent ferromagnetic response was found for all samples, consistent with the early experiments.[17,18] To achieve the insulating phase, a small amount of Bi was added to tune the chemical potential (see Supplementary Information).[20,21] We note that $T_C$ and $H_c$ varies little with Bi doping, even in the rather insulating samples around the p-n crossover region, which indicates that there is a truly insulating ferromagnetic state in V-doped $Sb_2Te_3$.[21] In order to realize accurate and continuous tuning of the Fermi energy and carrier density ($n_{2D}$), electric-field gating was also employed. The heat-treated $SrTiO_3(111)$ substrate with a large dielectric constant serves as an effective bottom gate dielectric. The magneto-transport was measured on two samples, denoted as S1 and S2, using the standard six-terminal Hall bar configuration. (see Methods and Supplementary Information).[14,21]

Our magneto-transport on a four quintuple-layers (QL) $(Bi_{0.29}Sb_{0.71})_{1.89}V_{0.11}Te_3$ film (sample S1) as shown in Fig. 1, exhibits nearly ideal QAH behavior. At the charge neutral point $V_g^0$, the Hall resistance ($\rho_{yx}$) at zero magnetic field (labeled as $\rho_{yx}(0)$) displays the quantized value of $1.00019\pm0.00069 h/e^2$ ($25.8178\pm0.0177 k\Omega$), while the longitudinal resistance ($\rho_{xx}$) at zero magnetic field (labeled as $\rho_{xx}(0)$) is only $0.00013\pm0.00007 h/e^2$ (~$3.35\pm1.76\Omega$) measured at $T$=25mK, shown in Figs 1b and 1c. The ratio $\rho_{yx}(0)/\rho_{xx}(0)$ corresponds to an anomalous Hall angle $\alpha$ of $89.993\pm0.004°$. The corresponding Hall conductance at zero magnetic field (labeled as $\sigma_{yx}(0)$) is found to be $0.9998\pm0.0006 e^2/h$ and the longitudinal conductance at zero magnetic field (labeled as $\sigma_{xx}(0)$), $0.00013\pm0.00007 e^2/h$. For comparison, $\rho_{xx}(0)$, $\sigma_{yx}(0)$, $\sigma_{xx}(0)$ and the anomalous Hall angle in the Cr-doped system are $0.098 h/e^2$(~$2.53k\Omega$), $0.987 e^2/h$, $0.096 e^2/h$ and $84.40°$ at $T$=30mK, respectively,[14] noticeably different from the expected values. The precise values of these quantities at zero-field were not reported systematically in ref. 15 and 16.

The temperature dependence of QAH behavior (sample S1) at the charge natural point $V_g^0$ after magnetic training was shown in Fig 2a. Below $T$=5K, $\rho_{yx}$ (same as $\rho_{yx}(0)$) shows a continuous increase and $\rho_{xx}$ (same as $\rho_{xx}(0)$) a continuous decrease with decreasing temperature, thus indicating that the QAH state in sample S1 survives up to ~5K. Figure 2b displays the temperature dependence of the anomalous Hall angle $\alpha$. The $\alpha$ grows rapidly with decreasing temperature, and asymptotically approaches the ideal value 90° at the lowest temperature ~25mK, which is consistent with previous Hall results. In the prior experiments in the Cr-doped system, the largest Hall angle was about $84.40°$ at the lowest temperature ~30mK.[14] In the V-doped system, an anomalous Hall angle of such a value is found near 130mK. (see Supplementary Information for a detailed comparison).

What we have found in V-doped system is clearly a more precise experimental confirmation of the ideal QAH effect than that in Cr-doped system. To understand the differences between these two materials, we carried out direct comparison of the two systems. Our studies showed three main differences. (1) Figure 3a shows the dependence of $T_C$ for 6QL films of $Sb_{2-x}Cr_xTe_3$ and $Sb_{2-x}V_xTe_3$ as a function of the Cr or V doping concentration $x$. The $T_C$ of a FM material can be determined from the Arrott plot (see Supplementary Information).[22] At the same concentration $x$, the $T_C$ for V doping is twice the value found for Cr doping. Having a higher $T_C$ with low $x$ is in the right direction for reaching the QAH state at a higher temperature.[23] (2) The Hall traces for the $x=0.13$ films are shown in Fig. 3b. In both cases, the square-shaped loops indicate the long-range ferromagnetic order with out-of-plane magnetic anisotropy.[24] The coercive field $H_c$ for the $Sb_{1.87}V_{0.13}Te_3$ film is ~1.3T at $T=2K$, an order of magnitude larger than that of $Sb_{1.87}Cr_{0.13}Te_3$ with $H_c$~0.1T.[14-16,21] The large $H_c$ of V-doped $Sb_2Te_3$ is clearly advantageous for a robust FMTI system.[25,26] (3) Remarkably, the $n_{2D}$ of 6QL $Sb_{1.87}V_{0.13}Te_3$ is ~ $2.5\times10^{13}cm^{-2}$, about one half of $n_{2D}$ in $Sb_{1.87}Cr_{0.13}Te_3$ ($n_{2D}$~$4.9\times10^{13}cm^{-2}$). The low $n_{2D}$ in the parent material is beneficial because it reduces the amount of Bi doping required to drive the system towards a charge neutral state.[20]

The main difference between these two materials lies in different magnetic moments and valence of V and Cr atoms. The saturation magnetic moment per V ion was determined to be 1.5$\mu_B$ (Bohr magnetron) for $x=0.13$ V-doped $Sb_{2-x}V_xTe_3$ (shown in Fig. S6). This suggests that the valence state of V is a mixture of 3+ and 4+ (or/and 5+) as it is expected to substitute for the $Sb^{3+}$ ion on the Sb sub-lattice.[17,18] The extra free electrons resulting from $V^{4+}$ (or/and $V^{5+}$) go to neutralize the $p$-type carriers, so the $n_{2D}$ of V-doped $Sb_2Te_3$ is expected and measured to be lower than that of Cr-doped samples, in which Cr takes the valence state 3+.[17,21]

The one order of magnitude higher $H_c$ found in the V-doped system is due to the much smaller magnetic moment of V atom as compared with Cr atom (~3$\mu_B$).[17,21] The $H_c$ in FM films, in addition to extrinsic influences such as defects that pin domain walls, depends on the magnetic anisotropy (both crystallographic and shape) and saturation magnetization moment ($M_s$) of the films. For magnetization reversal by the process of magnetic domain rotation, $H_c \propto \frac{K}{M_s}$, where $K$ is the total magnetic anisotropy constant, so a higher $K$ and/or lower $M_s$ can lead to a high $H_c$. Since the demagnetization energy ($E_d$) determines the number of magnetic domain walls, and $E_d \propto M_s^2$, materials with few domains have a high $H_c$ while those with many domains have a low $H_c$.[25,26] Therefore it is clear that V-doped $Sb_2Te_3$ with a lower $M_s$ and a higher $H_c$ have fewer magnetic domains than Cr-doped systems, indicating that the magnetic domains in V-doped $Sb_2Te_3$ are much larger in size than in Cr-doped films. The two main sources of magnetic anisotropy, magnetocrystalline anisotropy and surface anisotropy, appear to be very different for V- and Cr-doped $Sb_2Te_3$ films. The magnetocrystalline anisotropy is usually the most effective means of impeding the magnetization reversal by rotation processes, while the surface anisotropy is a major contributor in nm thick films. In high-quality epitaxial V-doped $Sb_2Te_3$ with large atomically flat areas, a near-perfect layered structure, and the slight expansion of $c$-axis lattice parameter may lead to the large magnetic anisotropy in the present films (see Supplementary Information).[25,26]

Due to the more robust ferromagnetism, V-doped $Sb_2Te_3$ films exhibit QAH state spontaneously without any magnetic field training, in sharp contrast to the Cr-doped systems.[14,15,16] Figure 4a shows the temperature-dependence of $\rho_{xx}$ and $\rho_{yx}$ for a different 4QL $(Bi_{0.29}Sb_{0.71})_{1.89}V_{0.11}Te_3$ film(sample S2), using a bottom gate bias $V_g=0$ and no applied magnetic field and cooled from room temperature under zero-field. $\rho_{xx}$ shows semiconducting behaviors from room temperature down to ~2.5K, where a drop occurs, signaling that it is entering the QAH state. $\rho_{yx}$ remains constant from room temperature down to ~23K where it increases rapidly with further decreasing in temperature. This indicates that the system enters the FM state below ~23K. $\rho_{yx}$ continues to increase and $\rho_{xx}$ continues to decrease with decreasing temperature. $\rho_{yx}$ approaches ~$0.99h/e^2$, and $\rho_{xx}$ reaches ~$0.30h/e^2$ at the lowest temperature of the measurement of sample S2 at ~80mK. This behavior signifies the film spontaneously progressing toward the QAH state. The net magnetization of virgin state in the film, *i.e.* not induced by any external field, and the reduced thermal activation at lower temperature drives the film from the diffusive transport regime towards the conducting QAH state. The inset of Fig. 4a shows the low-temperature data at $V_g=0V$ below 3K, for magnetic fields $\mu_0H=0$ and 2T. This spontaneous decrease of $\rho_{xx}$ at zero-field seen for V-doping is absent in Cr-doped TIs.[14,15,16] In Cr-doped TI films, unless the magnetic domains are aligned with an external applied perpendicular magnetic field, $\rho_{xx}$ increases with decreasing temperature and no QAH state can be achieved even at very low temperatures.[15,16] Applying a field of 2T to the V-doped TI film, on the other hand, did not change the qualitative temperature dependence in $\rho_{xx}$ and $\rho_{yx}$, from that of the zero-field cooling behavior. This clearly demonstrates that the macroscopic net magnetization of the virgin state in V-doped TIs is spontaneous and highly developed. The fact that virgin $\rho_{yx}$ and $\rho_{xx}$ curves overlap with the hysteresis loop (Fig. 4c) and the butterfly structure (Fig. 4d), also confirms the spontaneous self-magnetization of V-doped TI samples in the QAH regime. The high self-magnetization also appears in the direct magnetization measurements (see Supplementary Information).

The QAH state observed in V-doped TIs is unexpected from the view of first-principle calculations that points to the absence of an insulating FM state in V-doped TI due to the presence of *d*-orbital impurity bands in the bulk band gap.[12] A possible explanation why there exists an insulating FM state could be that the *d*-orbital impurity bands become localized at lower temperatures, so that it does not participate in the charge transport and the whole system becomes an Anderson insulator instead of a band insulator. The localization physics in the QAH regime has been studied numerically in Ref. 10, which confirms that the QAH state can indeed be achieved in a metallic system by introducing disorder to localize bulk carriers. Thus, our experiments indicate the role of the localization physics of impurity bands in V-doped systems.

This physical picture also provides a natural explanation of high $T_C$ observed in V-doped $Sb_2Te_3$, compared to that in Cr-doped $Sb_2Te_3$. The van Velck mechanism depends on the inverted band structure, and is similar for these two systems since their parent materials are the same. The impurity bands can mediate double exchange interaction between magnetic moments, which is well known for some diluted magnetic semiconductors, particularly in the insulating regime.[27] Thus, we expect that the double exchange interaction provides additional channels for ferromagnetism, leading to the enhancement of the $T_C$ in V-doped $Sb_2Te_3$. This explanation is further supported by the fact that the gate-voltage dependence of $H_c$ of V-doped film (see Supplementary Information) is much larger than that of Cr-doped system.[14] This scenario possibly paves a new path to enhance the observation temperature of the QAH effect.

Our results demonstrate the high-precision confirmation of the QAH state with a zero-field longitudinal resistance as low as $0.00013 \pm 0.00007 h/e^2$ (~$3.35 \pm 1.76\Omega$) and zero-field Hall conductance reaching ~0.9998±0.0006$e^2/h$ in V-doped (Bi,Sb)$_2$Te$_3$, a hard FMTI. These precise values firmly establish the equivalence of the quantum Hall (QH) state and the QAH state. It is particularly remarkable considering that mobility of the device measured near 80K is only ~130cm$^2$/Vs (see Supplementary Information), orders of magnitude smaller than that found in semiconductor hetero-structures with two dimensional electron gas (2DEG) that reached the QH state.[1,2] Compared with the QH state, the QAH state with enough precision in this system has better potential in the applications of metrology. This FMTI is found to be in the QAH state without the aid of a polarizing external magnetic field, making it also a promising candidate for dissipationless electronic applications.

## Methods

**MBE growth**. Thin film growth was performed using a custom-built ultrahigh vacuum MBE system. Semi-insulating etched Si(111), insulating heat-treated sapphire(0001) and SrTiO$_3$(111) substrates were outgassed before the growth of TI films. High-purity Bi(99.999%), Sb(99.9999%), and Te(99.9999%) were evaporated from Knudsen effusion cells, whereas the transition metal dopants Cr(99.999%) and V(99.995%) were evaporated by e-guns. During the growth, the substrate was maintained at 230℃. The flux ratio of Te per Bi and Sb was set to approximately ~8 to prevent Te deficiency in the films. The Sb, Bi, and V concentration in the films were determined by their ratio obtained in situ during growth using separate quartz crystal monitors and later confirmed ex situ by inductively coupled plasma atomic emission spectroscopy (ICP-AES). The growth rate for the films was approximately 0.2 quintuple layers per minute. Epitaxial growth was monitored by in situ reflection high energy electron diffraction (RHEED) patterns, where the high crystal quality and the atomically flat surface were confirmed by the streaky and sharp "1×1" patterns (see Supplementary Information).

**Transport measurements.** The transport measurements were performed ex situ on the magnetically-doped TI thin films. To avoid possible contamination, a 10-nm thick epitaxial Te capping layer was deposited at room temperature on top of the TI films before taken out of the growth chamber for transport measurements (see Supplementary Information). The Hall effect and longitudinal resistance were measured using both a Quantum Design Physical Property Measurement System (PPMS) (50mK, 9T) and a dilution refrigerator (Leiden Cryogenics, 10mK, 9T) with the excitation current flowing in the film plane and the magnetic field applied perpendicular to the plane. The bottom gate voltage was applied using the Keithley 6430. All the resistance meters were calibrated by a standard resistor. The QAH results reported here have been reproduced on several samples using the above two cryostats. $\rho_{yx}$ and $\rho_{xx}$ results shown in Fig. 1 were carried out with excitation current of 1 nA and with the mixing chamber of the dilution refrigerator anchored at 10 mK. However, systematic measurements of $\rho_{xx}$ as functions of temperature and (higher) excitation current indicate the electron temperature of the sample is ~25±5mK due to the stray electromagnetic noise.

# References

1. Klitzing, K. v., Dorda, G. & Pepper, M. New method for high-accuracy determination of the fine-structure constant based on quantized Hall resistance. *Phys. Rev. Lett.* **45**, 494-467 (1980).

2. Beenakker, C.W.J. and Houten, H. van. Quantum transport in semiconductor nanostructures. *Solid State Phys.* **44**, 1-228 (1991).

3. Haldane, F. D. M. Model for a quantum Hall effect without Landau levels: condensed-matter realization of the 'parity anomaly'. *Phys. Rev. Lett.* **61**, 2015-2018 (1988).

4. Onoda, M. and Nagaosa, N. Quantized anomalous Hall effect in two-dimensional ferromagnets: quantum Hall effect in metals. *Phys. Rev. Lett.* **90**, 206601 (2003).

5. Liu, C. X. *et al*. Quantum anomalous Hall effect in $Hg_{1-y}Mn_yTe$ quantum wells. *Phys. Rev. Lett.* **101**, 146802 (2008).

6. Qiao, Z. H. *et al*. Quantum anomalous Hall effect in graphene proximity coupled to an antiferromagnetic insulator. *Phys. Rev. Lett.* **112**, 116404 (2014).

7. Qiao, Z. H. *et al*. Quantum anomalous Hall effect in graphene from Rashba and exchange effects. *Phys. Rev. B* **82**, 161414(R) (2010).

8. Zhang, H. B. *et al*. Electrically tunable quantum anomalous Hall effect in graphene decorated by *5d* transition-metal adatoms. *Phys. Rev. Lett.* **108**, 056802 (2012).

9. Ezawa, M. Valley-polarized metals and quantum anomalous Hall effect in silicene. *Phys. Rev. Lett.* **109**, 055502 (2012).

10. Nomura, K. *et al*. Surface-quantized anomalous Hall current and the magnetoelectric effect in magnetically disordered topological insulators. *Phys. Rev. Lett.* **106**, 166802 (2011).

11. Garrity, K. F. and Vanderbilt, D. Chern Insulators from Heavy Atoms on Magnetic Substrates. *Phys. Rev. Lett.* **110**, 116802 (2013).

12. Yu, R. *et al*. Quantized anomalous Hall effect in magnetic topological insulators. *Science* **329**, 61-64 (2010).

13. Qi, X. L., Hughes, T. L. & Zhang, S. C. Topological field theory of time-reversal invariant insulators. *Phys. Rev. B* **78**, 195424 (2008).

14. Chang, C. Z. *et al*. Experimental observation of the quantum anomalous Hall effect in a magnetic topological insulator. *Science* **340**, 167-170 (2013).

15. Kou, X. *et al*. Scale-invariant quantum anomalous Hall effect in magnetic topological insulators beyond two-dimensional limit. *Phys. Rev. Lett.* **113**, 137201 (2014).

16. Checkelsky, J. G. *et al*. Trajectory of the anomalous Hall effect toward the quantized state in a ferromagnetic topological insulator. *Nat. Phys.* **10**, 731(2014).

17. Chien, Y. J. Transition metal-doped $Sb_2Te_3$ and $Bi_2Te_3$ Diluted Magnetic Semiconductors. Ph.D. Dissertation. The University of Michigan (2007).

18. Dyck, J. S. *et al*. Diluted magnetic semiconductors based on $Sb_{2-x}V_xTe_3$ ($0.01 \leq x \leq 0.03$). *Phys. Rev. B* **65**, 115212 (2002).

19. Hor, Y. S. *et al*. Development of ferromagnetism in the doped topological insulator $Bi_{2-x}Mn_xTe_3$. *Phys. Rev. B* **81**, 195203 (2010).

20. Zhang, J. *et al*. Band structure engineering in $(Bi_{1-x}Sb_x)_2Te_3$ ternary topological insulators. *Nat. Commun.* **2**, 574 (2011).

21. Chang, C. Z. *et al*. Thin films of magnetically doped topological insulator with carrier-independent long-range ferromagnetic order. *Adv. Mater.* **25**, 1065-1070 (2013).

22. Arrott, A. Criterion for Ferromagnetism from Observations of Magnetic Isotherms. *Phys. Rev. B* **108**, 1394 (1957).

23. He, K. *et al*. Quantum anomalous Hall effect. *Natl. Sci. Rev.* **1**, 39-49 (2014).

24. Nagaosa, N. *et al.* Anomalous Hall effect. *Rev. Mod. Phys.* **82**, 1539-1592 (2010).

25. Coey, J. M. D. Magnetism and magnetic materials. Cambridge University Press, New York (2009).

26. Livingston, J. D. A review of coercivity mechanisms. *J. Appl. Phys.* **52(3)**, 2544-2548 (1981).

27. Coey, J. M. D. et al. Donor impurity band exchange in dilute ferromagnetic oxides. *Nat. Mater.* **4**, 173-179 (2005).

## Acknowledgements

We are grateful to P. Wei, J. Liu, L. Fu, N. Samarth, J. Jain and G. Csathy and Z. Fang for helpful discussions, and F. Katmis, W. J. Fang, C. Settens and J. Kong for technical support in characterizing the samples. This research is supported the grants NSF (DMR-1207469), NSF (DMR-0907007), ONR (N00014-13-1-0301), NSF (DMR-0820404, DMR-1420620, Penn State MRSEC), NSF (DMR-1103159), DOE (DE-AC02-76SF00515), DARPA (N66001-11-1-4105) and the STC Center for Integrated Quantum Materials under NSF grant DMR-1231319.

## Author contributions

C. Z. C., M. H. W. C. and J. S. M. conceived and designed the research. C. Z. C. grew the material with the help of J. S. M.. C. Z. C. performed characterization studies of the samples with the help of B. A. A. and D. H.. W. Z. made the devices and performed the transport measurements with the help of C. Z. C., D.

K. and M. H. W. C.. C. X. L., H. J. Z. and S. C. Z. provided theory support. C. Z. C., C. X. L., and M. H. W. C. analyzed the data and wrote the manuscript with contributions from all authors.

## Additional information

Supplementary information is available in the online version of the paper. Reprients and permissions information is available at www.nature.com/reprints.

Correspondences and requests for materials should be addressed to C. Z. C., W. Z. or J. S. M..

## Competing financial interests

The authors declare no competing financial interests.

**Figures and Figure captions:**

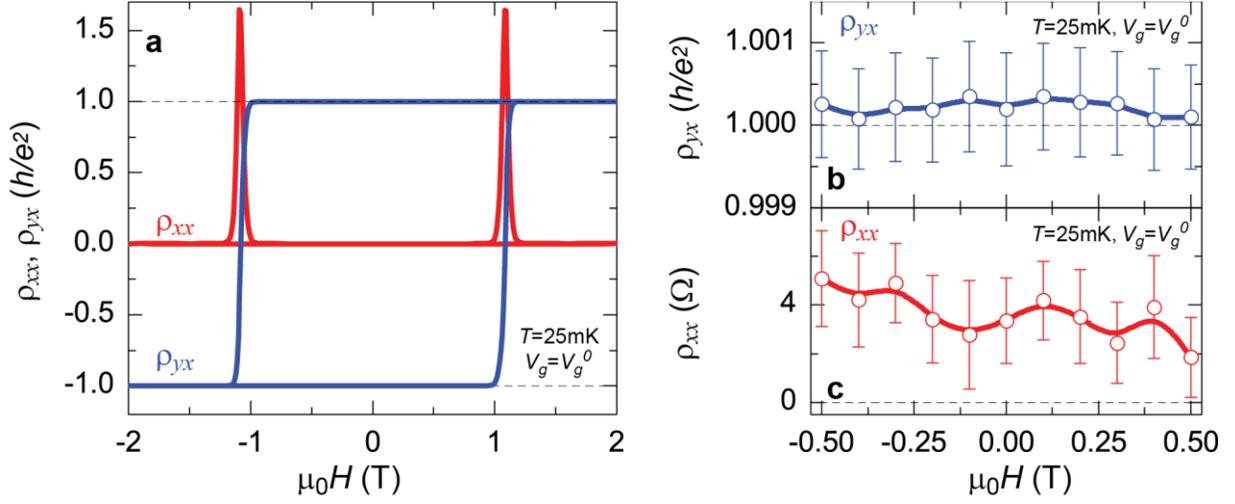

**Figure 1 | The QAH effect in a 4QL $(Bi_{0.29}Sb_{0.71})_{1.89}V_{0.11}Te_3$ film (sample S1) measured at 25mK. a,** Magnetic field dependence of the longitudinal resistance $\rho_{xx}$ (red curve) and the Hall resistance $\rho_{yx}$ (blue curve) at charge neutral point $V_g=V_g^0$. **b, c,** Expanded $\rho_{yx}$ (**b**) and $\rho_{xx}$ (**c**) at low magnetic field. $\rho_{yx}$ at zero magnetic field exhibits a value of $1.00019\pm0.00069 h/e^2$, while $\rho_{xx}$ at zero magnetic field is only ~$0.00013\pm0.00007 h/e^2$ (~$3.35\pm1.76\Omega$). The dashed lines indicate the expected resistance value in ideal QAH regime. The error bars in (**b**) and (**c**) are estimated by the variability of the standard resistor, accuracy of the voltmeters, current source which is reflected in the averaged data.

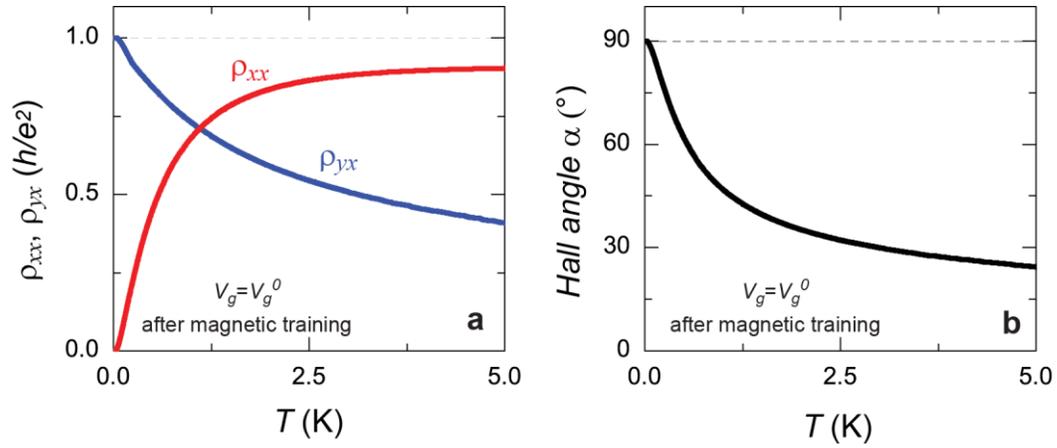

**Figure 2 | The temperature dependence of QAH behavior in a 4QL $(Bi_{0.29}Sb_{0.71})_{1.89}V_{0.11}Te_3$ film (sample S1). a**, Temperature dependence of the longitudinal resistance $\rho_{xx}$ (red curve) and Hall resistance $\rho_{yx}$ (blue curve) of sample S1 after magnetic training. **b**, Temperature dependence of the anomalous Hall angle $\alpha$, also shown is the ideal value (90°, dashed line). The dashed lines indicate the expected values in ideal QAH regime.

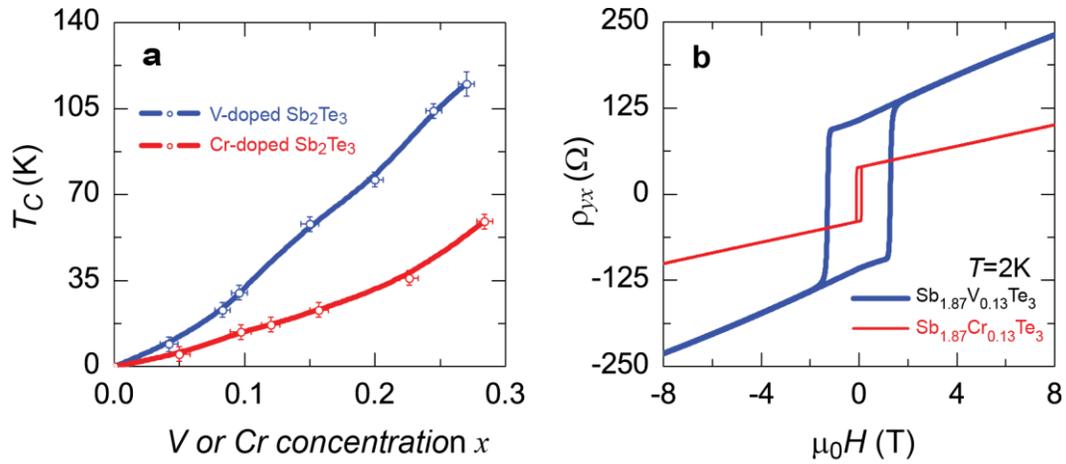

**Figure 3 | Ferromagnetic properties comparing Cr- and V-doped $Sb_2Te_3$. a**, The Curie temperature ($T_C$) of 6QL $Sb_{2-x}V_xTe_3$ (blue circles) and $Sb_{2-x}Cr_xTe_3$ thin films (red circles). Note the factor of 2 increase in $T_C$ for V doping. **b**, The Hall traces of 6QL $Sb_{1.87}V_{0.13}Te_3$ and $Sb_{1.87}Cr_{0.13}Te_3$ thin films, measured at $T$=2K. Note the order of magnitude increase in $H_c$ for V doping. The blue and red solid lines in (**a**) are guides to the eye. The horizontal and vertical error bars in (**a**) are estimated from the actual V or Cr content as opposed to the required amount during the film growth and the $T_C$ determination accuracy, respectively.

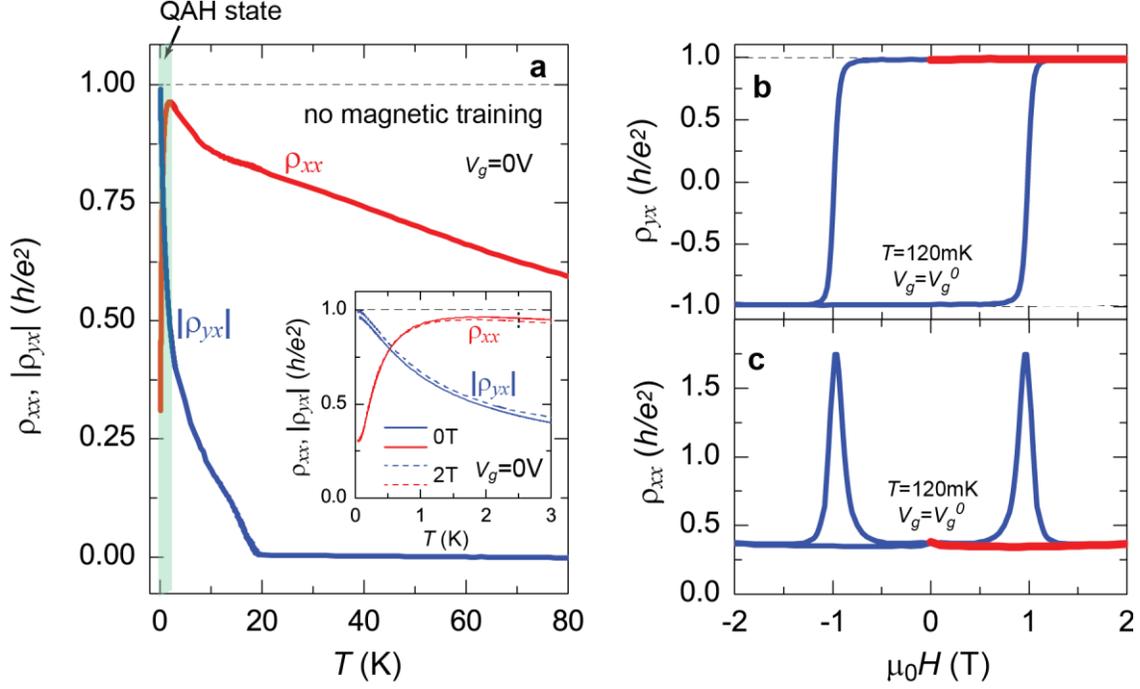

**Figure 4 | The self-driven QAH state in a different 4QL $(Bi_{0.29}Sb_{0.71})_{1.89}V_{0.11}Te_3$ film (sample S2). a**, Temperature dependence of longitudinal resistance $\rho_{xx}$ (red curve) and Hall resistance $|\rho_{yx}|$ (blue curve) of sample S2 without magnetic training. Inset: *T*-dependent $\rho_{xx}$ (red curve) and $|\rho_{yx}|$ (blue curve) under magnetic field $\mu_0 H=0$ (solid) and 2T (dashed). The opposite *T*-dependent behaviors of $\rho_{xx}$ and $|\rho_{yx}|$ reveal the QAH state at low *T*. **b, c**, The magnetic field-dependent $\rho_{yx}$ (**b**) and $\rho_{xx}$ (**c**) of sample S2 measured at 120mK. The red curves in **b** and **c** were obtained by first cooling the sample under zero-field and then measured from zero to higher field. The black dashed lines indicate the expected values in ideal QAH regime.